\documentclass{article}

 \usepackage[preprint]{neurips_2025}

\usepackage[utf8]{inputenc} 
\usepackage{amsmath}
\usepackage{makecell}
\usepackage{graphicx}
\usepackage{cellspace}
\usepackage{subcaption}
\usepackage{array}
\usepackage[T1]{fontenc}    
\usepackage{hyperref}       
\usepackage{url}            
\usepackage{booktabs}       
\usepackage{amsfonts}       
\usepackage{nicefrac}       
\usepackage{microtype}      
\usepackage{xcolor}         
\usepackage{tabularx}

\title{Mix-Geneformer: Unified Representation Learning for Human and Mouse scRNA-seq Data}

\author{%
  Yuki Nishio\\
  Department of Robotic Science and Technology\\
  Chubu University\\
  Kasugai Japan \\
  \texttt{yuki@mprg.cs.chubu.ac.jp}
  \And
  Prof. Takayoshi Yamashita\\
  Department of Computer Science\\
  Chubu University\\
  Kasugai Japan \\
  \texttt{takayoshi@isc.chubu.ac.jp}
  \And
  Keita Ito\\
  Department of Computer Science\\
  Chubu University\\
  Kasugai Japan \\
  \And
  Tsubasa Hirakawa\\
  Department of Center for Mathematical Science and Artificial Intelligence\\
  Chubu University\\
  Kasugai Japan\\
  \texttt{hirakawa@mprg.cs.chubu.ac.jp}
  \And
  Prof. Hironobu Fujiyoshi\\
  Department of Robotic Science and Technology\\
  Chubu University\\
  Kasugai Japan\\
  \texttt{fujiyoshi@isc.chubu.ac.jp}}
\begin{document}

\maketitle
\begin{abstract}

Single-cell RNA sequencing (scRNA-seq) enables single-cell transcriptomic profiling, revealing cellular heterogeneity and rare populations. Recent deep learning models like Geneformer and Mouse-Geneformer perform well on tasks such as cell-type classification and in silico perturbation. However, their species-specific design limits cross-species generalization and translational applications, which are crucial for advancing translational research and drug discovery. We present Mix-Geneformer, a novel Transformer-based model that integrates human and mouse scRNA-seq data into a unified representation via a hybrid self-supervised approach combining Masked Language Modeling (MLM) and SimCSE-based contrastive loss to capture both shared and species-specific gene patterns. A rank-value encoding scheme further emphasizes high-variance gene signals during training. Trained on about 50 million cells from diverse human and mouse organs, Mix-Geneformer matched or outperformed state-of-the-art baselines in cell-type classification and in silico perturbation tasks, achieving 95.8\% accuracy on mouse kidney data versus 94.9\% from the best existing model. It also successfully identified key regulatory genes validated by in vivo studies. By enabling scalable cross-species transcriptomic modeling, Mix-Geneformer offers a powerful tool for comparative transcriptomics and translational applications. While our results demonstrate strong performance, we also acknowledge limitations, such as the computational cost and variability in zero-shot transfer.

\end{abstract}

\section{Introduction}
Single-cell RNA sequencing (scRNA-seq) analysis~\cite{scRNA} is a powerful technique that enables the acquisition of gene expression information at the individual cell level, providing unprecedented insights into cellular heterogeneity and rare cell populations. This technology has become indispensable for understanding complex biological processes, including cancer and immune responses, and has generated vast datasets across diverse organisms, such as humans and mice.  The increasing availability of these data necessitates the development of advanced analytical methods to fully leverage their potential for driving progress in medicine and the life sciences. 

Deep learning models, particularly those based on the Transformer architecture~\cite{transformer}, have shown remarkable success in analyzing large-scale and high-dimensional scRNA-seq data.  Models like Geneformer~\cite{geneformer}  and Mouse-Geneformer~\cite{mouse-geneformer} treat gene expression profiles as "sentences" and genes as "words," using attention mechanisms to capture intricate contextual relationships between genes. These models have proven valuable for various downstream tasks, including cell-type classification and in silico gene perturbation simulations, offering promising applications in drug discovery and regenerative medicine.  

However, Geneformer is specifically trained on human scRNA-seq data, limiting its generalization to other species. While Mouse-Geneformer  addresses this by training on mouse data and achieving high performance in mouse-specific tasks, both models are species-specific and cannot effectively handle data from both humans and mice simultaneously.  This limitation hinders their ability to robustly handle domain shifts between species and restricts their translational potential. 

To address this critical need, we introduce Mix-Geneformer, a novel Transformer-based model designed for joint learning from human and mouse scRNA-seq data to enable effective cross-species analysis.  Mix-Geneformer employs a hybrid self-supervised learning strategy, combining Masked Language Modeling (MLM) with SimCSE-based contrastive learning, to learn consistent gene representations across species.  This approach enhances analytical efficiency and facilitates real-world biological research.   

\paragraph{Contributions.}
The main contributions of this work are:
\begin{itemize}
  \item We propose Mix-Geneformer, a Transformer-based model for joint learning of scRNA-seq data from human and mouse.
  \item We introduce a hybrid self-supervised learning scheme, integrating Masked Language Modeling (MLM) and SimCSE-based contrastive learning, to achieve consistent gene representations across species.  
  \item We demonstrate that Mix-Geneformer achieves competitive or superior performance compared to species-specific models in cell-type classification and in silico perturbation tasks, using a large-scale dataset of approximately 50,000,000 cells from human and mouse.  
  \item We evaluate zero-shot transfer capabilities in both human → mouse and mouse → human directions, showing the utility of our approach for translational research in drug discovery and disease studies.  
\end{itemize}

\section{Related Work}
Single-cell RNA sequencing (scRNA-seq) has revolutionized the study of gene expression by enabling the acquisition of transcriptomic information at the individual cell level. scRNA-seq offers a significant advantage over traditional bulk RNA-seq methods. This technology reveals cellular heterogeneity and allows for the characterization of rare cell types that are often masked in population-averaged data~\cite{scRNA}. Consequently, scRNA-seq data have become an indispensable resource for investigating diverse biological processes, including cancer development and immune responses, and are now widely available for a broad range of organisms, such as humans and mice.


The analysis of large-scale, high-dimensional scRNA-seq datasets has greatly benefited from advancements in deep learning. Various deep learning architectures have been applied to scRNA-seq data for tasks such as dimensionality reduction and imputation~\cite{scvae}. In particular, Transformer-based models have emerged as powerful tools for capturing complex relationships within gene expression profiles~\cite{enformer}.

The Transformer architecture~\cite{transformer}, originally developed for natural language processing, has proven to be highly effective in modeling gene–gene interactions in scRNA-seq data. These models treat gene expression profiles as “sentences” and individual genes as “words,” using attention mechanisms to learn contextual relationships between genes. Geneformer~\cite{geneformer} is a prominent example, pre-trained on large-scale human scRNA-seq data using Masked Language Modeling (MLM) and demonstrating strong performance in human-specific downstream tasks, including cell-type classification and in silico perturbation experiments. Mouse-Geneformer \cite{mouse-geneformer} adopts the same Transformer architecture and pre-training strategy as Geneformer but is trained on mouse scRNA-seq data, achieving high accuracy in mouse-specific analyses.  
Additionally, scDeepSort~\cite{scdeepsort} employs a weighted graph neural network (GNN) framework to perform cell-type annotation on single-cell transcriptomics. It was pre-trained on two high-quality scRNA-seq atlases comprising 764,741 cells across 88 human and mouse tissues, achieving an overall annotation accuracy of 83.79\% on 76 external test datasets without reliance on marker genes or reference profiles~\cite{scdeepsort}.

Self-supervised learning techniques, such as Masked Language Modeling (MLM)~\cite{bert}, have been crucial for pre-training Transformer models on unlabeled scRNA-seq data. MLM enables the model to learn context-dependent representations of genes by predicting masked gene expression values, effectively capturing gene co-expression patterns. scBERT~\cite{scbert} is another example of a Transformer-based model for single-cell analysis that utilizes MLM for pre-training, achieving impressive results in cell-type annotation and batch effect correction. MLM has also been successfully applied in genomics, as demonstrated by DNABERT~\cite{dnabert}, which is pre-trained on DNA sequences for various genomic analyses.

Pre-trained models like Geneformer and Mouse-Geneformer can be effectively utilized for various downstream tasks through transfer learning. By fine-tuning these models on specific tasks, such as cell-type classification or in silico perturbation, researchers can leverage the learned gene representations to gain insights into biological processes and disease mechanisms. Transfer learning has also been shown to be a valuable approach in network biology, as demonstrated in~\cite{geneformer}.

Despite their success, Geneformer and Mouse-Geneformer are inherently species-specific, limiting their ability to directly analyze and compare scRNA-seq data from different species. This limitation hinders cross-species generalization and translational applications, which are crucial for translating findings from model organisms (e.g., mice) to human studies. While human and mouse gene expression patterns share conserved features, they also exhibit significant species-specific regulatory mechanisms and evolutionary differences. Consequently, existing species-specific models struggle to capture both shared and unique features, limiting their effectiveness in integrated cross-species analysis. To address this challenge and accelerate translational research, there is a growing need for general-purpose models capable of jointly analyzing scRNA-seq data from multiple species.



\section{Proposed Method}

Geneformer and Mouse-Geneformer are Transformer models trained on human and mouse scRNA-seq data, showing strong performance in cell-type classification and in silico perturbation within their respective species. However, these models are species-specific, and a model capable of cross-species data handling has not yet been developed. Consequently, existing models show limited robustness to domain shifts between species, reducing their transferability~\cite{mouse-geneformer}. While human and mouse expression share conserved structures, they also exhibit species-specific mechanisms, making it difficult for current models to capture both human and mice. Biological research necessitates translating mouse findings to human systems, creating a demand for a general model for cross-species gene expression analysis.

To meet this requirement, this study constructs a general-purpose Transformer model for multiple species by training on human and mouse scRNA-seq data using Masked Language Modeling (MLM), a self-supervised learning method also used in scBERT and DNABERT. This integrated approach allows visualization of human-mouse relationships in feature space, potentially revealing novel cross-species gene networks and functional associations. This is expected to enable various cross-species applications in basic biology, disease analysis, and drug discovery. The following sections detail Mix-Geneformer's architecture, the datasets, training, and potential applications.

\subsection{Architecture of Mix-Geneformer}
\begin{figure}[tb]
    \centering
    \includegraphics[width=0.9\linewidth]{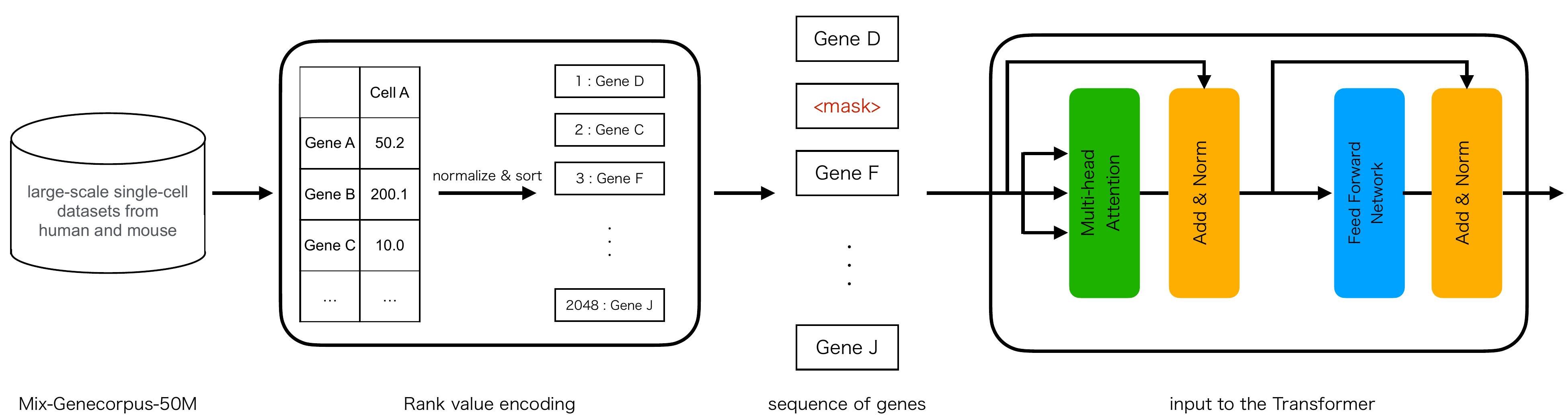}
    \caption{Overall architecture of Mix-Geneformer}
    \label{mix-geneformer}
\end{figure}

Mix-Geneformer adopts a BERT-based architecture, similar to Geneformer and Mouse-Geneformer, to effectively capture complex gene-gene relationships. The model consists of six encoder layers and four attention heads.  

\textbf{Transformer Encoder}: The Transformer encoder is a neural network architecture that relies entirely on attention mechanisms to draw global relationships between input elements. In Mix-Geneformer, the input is a sequence of gene expression values for a given cell, and the attention mechanism allows the model to weigh the importance of each gene in the context of all other genes in that cell. This is crucial for capturing synergistic and regulatory effects between genes.

\textbf{BERT-based Architecture}: BERT is a Transformer-based model pre-trained on large text corpora to learn contextual word representations. Geneformer and Mouse-Geneformer have demonstrated the effectiveness of applying the BERT architecture to scRNA-seq data, treating genes as "words" and cells as "sentences." Mix-Geneformer extends this approach to the cross-species setting.

Figure\ref{mix-geneformer} provides an overview of the Mix-Geneformer architecture. The key architectural difference between Mix-Geneformer and its predecessors lies in its joint training on human and mouse data and the incorporation of contrastive learning, in addition to Masked Language Modeling (MLM), to facilitate the learning of consistent gene representations across species.


\subsection{Pre-training Dataset}
To pre-train Mix-Geneformer, we constructed Mix-Genecorpus-50M, a large-scale combined dataset of human and mouse scRNA-seq data. This dataset integrates Genecorpus-30M, the pre-training dataset for Geneformer (human data), and Mouse-Genecorpus-20M, the pre-training dataset for Mouse-Geneformer (mouse data). 

\textbf{Data Quality Control}: Both Genecorpus-30M and Mouse-Genecorpus-20M were carefully curated to include only high-quality single-cell data.  Cells with evidence of non-cellular RNA, cell doublets, or low viability were excluded to minimize noise and artifacts in the pre-training data. 

\textbf{Rank Value Encoding}: Following data filtering, we applied rank value encoding, a crucial preprocessing step, to the gene expression values.  This encoding scheme transforms the raw expression values into a ranked representation, emphasizing the relative expression levels of genes within each cell.  The procedure involves: 

1. Normalization: Dividing each gene's expression value by the total expression within the cell. 

2. Scaling: Multiplying the normalized values by the median of non-zero expression values and then by 10,000. 

3. Feature Selection: Extracting the top 2,048 genes with the highest normalized scores.

Rank value encoding has been shown to improve the ability of Transformer models to learn biologically meaningful gene patterns from scRNA-seq data by focusing on the most informative, high-variance genes that characterize cell states.

\subsection{Pre-training}
Mix-Geneformer is pre-trained using a hybrid self-supervised learning objective that combines Masked Language Modeling (MLM) loss and SimCSE loss.
The total loss $\mathcal{L}_{\mathrm{total}}$ is defined as follows:
 \begin{align}
 \mathcal{L}_{\mathrm{total}} &= \mathcal{L}_{\mathrm{MLM}} + \mathcal{L}_{\mathrm{SimCSE}},
 \end{align}
 Here, $\mathcal{L}{\mathrm{MLM}}$ denotes the MLM loss, and $\mathcal{L}{\mathrm{SimCSE}}$ denotes the SimCSE loss. 

MLM is a self-supervised technique where a certain percentage of input tokens are randomly masked, and the model is trained to predict the original tokens. This forces the model to learn contextual relationships between genes, as it must rely on the surrounding genes to infer the masked gene's expression. The MLM loss is calculated as: 
  \begin{align}
     \mathcal{L}_{\mathrm{MLM}} &= - \sum_{i \in \mathcal{M}} \log P(x_i \mid x_{\setminus \mathcal{M}}; \theta),
 \end{align}
 Here, $\mathcal{M}$ denotes the set of masked token positions, $x_i$ is the target token, $x_{\setminus \mathcal{M}}$ is the unmasked token sequence, and $\theta$ represents the model parameters. 
 
Contrastive learning aims to learn representations where similar data points are close to each other in the embedding space, while dissimilar points are far apart. SimCSE is a contrastive learning method that uses slightly perturbed versions of the same input as positive pairs. In our case, two slightly different "views" of the same cell's gene expression profile are created, and the model learns to produce similar representations for these views. 
The SimCSE loss is defined as:
 \begin{align}
     \mathcal{L}_{\mathrm{SimCSE}} &= - \sum_{i=1}^{N} \log \frac{\exp(\mathrm{sim}(h_i, h_i^+)/\tau)}{\sum_{j=1}^{2N}  \mathbf{1}_{[\,j \neq i\,]}  \exp(\mathrm{sim}(h_i, h_j)/\tau)},
 \end{align}
The combination of MLM and SimCSE loss encourages the model to learn both context-dependent gene relationships (MLM) and semantically consistent cell representations across species (SimCSE).
 
\par In the pretraining, we used a maximum input sequence length of 2,048, six Transformer encoder layers, four attention heads, an embedding dimension of 256, the SiLU activation function, a maximum learning rate of 1e-4 (with a cosine scheduler), a batch size of 8 per GPU, and trained for 10 epochs. Through this pretraining, the model learns the relationships between human and mouse genes, enabling potential applications to various downstream tasks.

\subsection{Potential Applications}

Mix-Geneformer's ability to jointly analyze human and mouse scRNA-seq data opens up several exciting possibilities for advancing biological research and translational medicine. 

\textbf{Accelerated Translational Research}: The traditional approach of validating findings from mouse models in human studies is often time-consuming and resource-intensive, requiring separate analyses and manual comparisons. Mix-Geneformer can streamline this process by enabling integrated analysis, facilitating the identification of conserved and species-specific gene expression patterns. This can accelerate the translation of mouse model findings to human applications, reducing the time and cost of drug development and disease studies.

\textbf{Improved Disease Modeling}: By learning a shared representation space for human and mouse cells, Mix-Geneformer can facilitate the identification of more accurate and relevant mouse models for human diseases. This can lead to more effective preclinical studies and a better understanding of disease mechanisms.

\textbf{Discovery of Cross-Species Gene Networks}: Mix-Geneformer can be used to uncover novel cross-species gene regulatory networks and functional associations that may be missed by species-specific analyses. This can provide new insights into fundamental biological processes and disease etiology.

\textbf{Drug Discovery}: The model can aid in the identification of potential drug targets that are conserved across species, increasing the likelihood of successful translation from animal models to humans. It can also be used to predict drug efficacy and toxicity in different species.

\section{Experiments}
\label{sec:experiments}

This section details the experiments conducted to evaluate Mix-Geneformer's performance and potential for cross-species analysis. We assessed the model's capabilities in two key tasks: cell-type classification and in silico perturbation.

\subsection{Experimental Overview}
We evaluated Mix-Geneformer through cell-type classification and in silico perturbation experiments. In Cell-type classification experiments, We compared Mix-Geneformer's classification accuracy against existing methods using scRNA-seq data from various human and mouse organs. We also evaluate the impact of SimCSE-based contrastive learning by comparing models trained with and without this component. As in silico perturbation experiments, We investigated Mix-Geneformer's ability to identify genes relevant to specific disease states by simulating gene perturbations and comparing the results to findings from in vivo studies. In vivo studies are experiments performed within living organisms to observe the effects of treatments or genetic modifications.

\subsection{Overview of In Silico Perturbation Experiments}
The similarity between cell states before and after perturbation is quantitatively assessed using the cosine similarity between their embedding representations.
In silico perturbation experiments computationally simulate genetic modifications, such as gene deletions or activations, to analyze gene function and predict their effects on gene networks.
To simulate gene deletion, the target gene's rank value was decreased, while the rank values of other genes were relatively increased. To simulate gene activation, the target gene's rank value was increased, while other genes' rank values were relatively decreased. The effect of the perturbation is measured by assessing the similarity between cell states before and after the perturbation. If the perturbed cell state becomes more similar to a specific target cell state, the perturbed gene is inferred to play a significant role in driving that state transition. Similarity was quantitatively measured using cosine similarity between cell embedding representations.

\subsection{Experimental Conditions}

We conducted cell-type classification experiments, fine-tuning both Mix-Geneformer and baseline models on the cell-type classification task before evaluation. The dataset comprised scRNA-seq data from various human and mouse organs.The human data includes cells from nine organs (spleen, brain, immune, kidney, large intestine, liver, lung, pancreas, and placenta).
The mouse data includes cells from nine organs (brain, heart, kidney, large intestine, limb muscle, mammary gland, spleen, thymus, and tongue). Each organ dataset was randomly split into training and test sets (80:20 ratio).
For fine-tuning, we used a maximum input length of 2048, 10 training epochs, a peak learning rate of 5e-5, a batch size of 8, 500 warm-up steps, and the AdamW optimizer. Training was performed using eight A100 GPUs over the course of approximately three days.

\subsection{Evaluation by Cell-Type Classification}
We evaluated cell-type classification performance using both fine-tuned models and in a zero-shot setting.

\subsubsection{Cell-Type Classification in Mouse}
\begin{figure}[tb]
    \centering
    \begin{subfigure}[b]{0.48\linewidth}
        \centering
        \includegraphics[width=\linewidth]{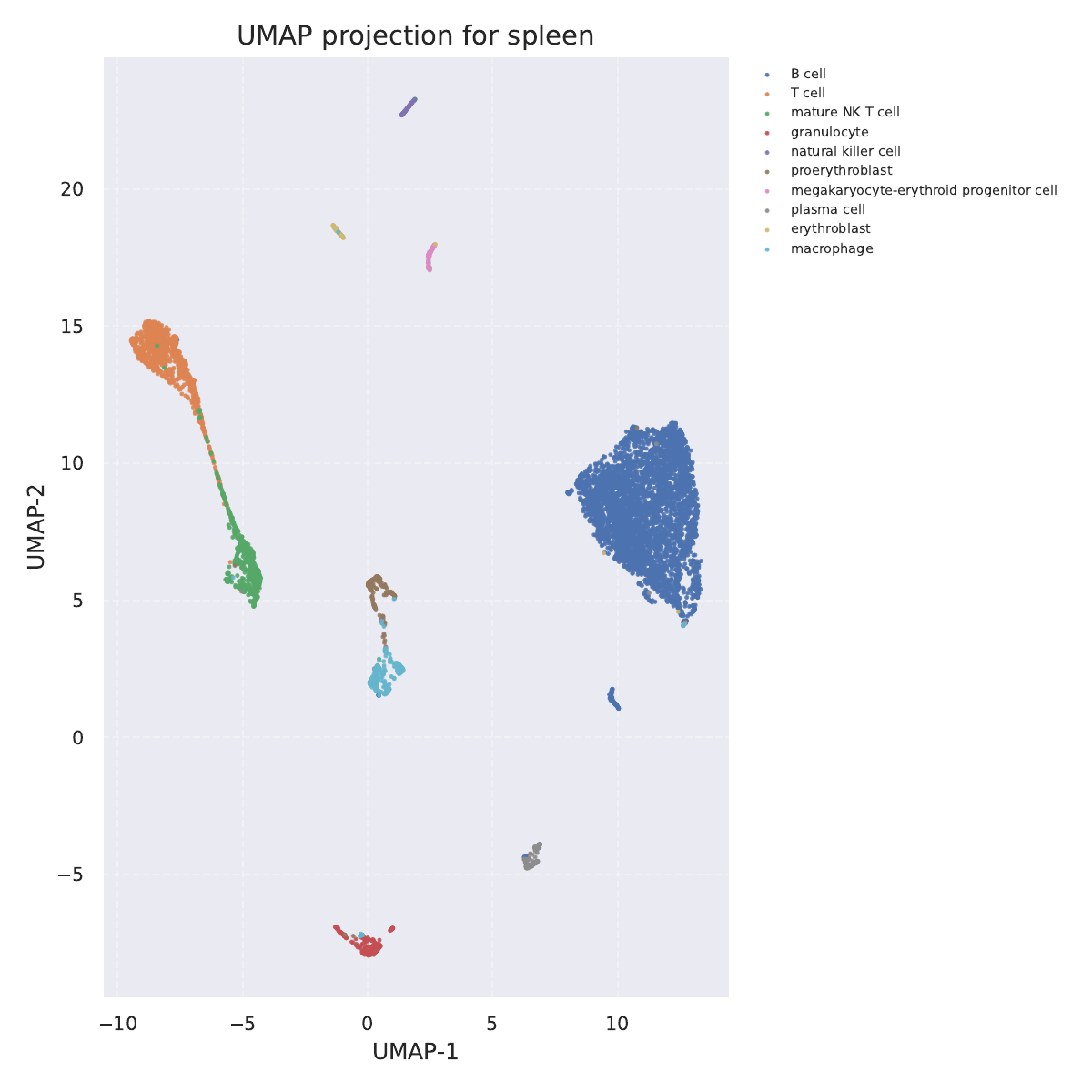}
        \caption{Mix-Geneformer}
        \label{fig:umap_mix_mouse}
    \end{subfigure}
    \hfill
    \begin{subfigure}[b]{0.48\linewidth}
        \centering
        \includegraphics[width=\linewidth]{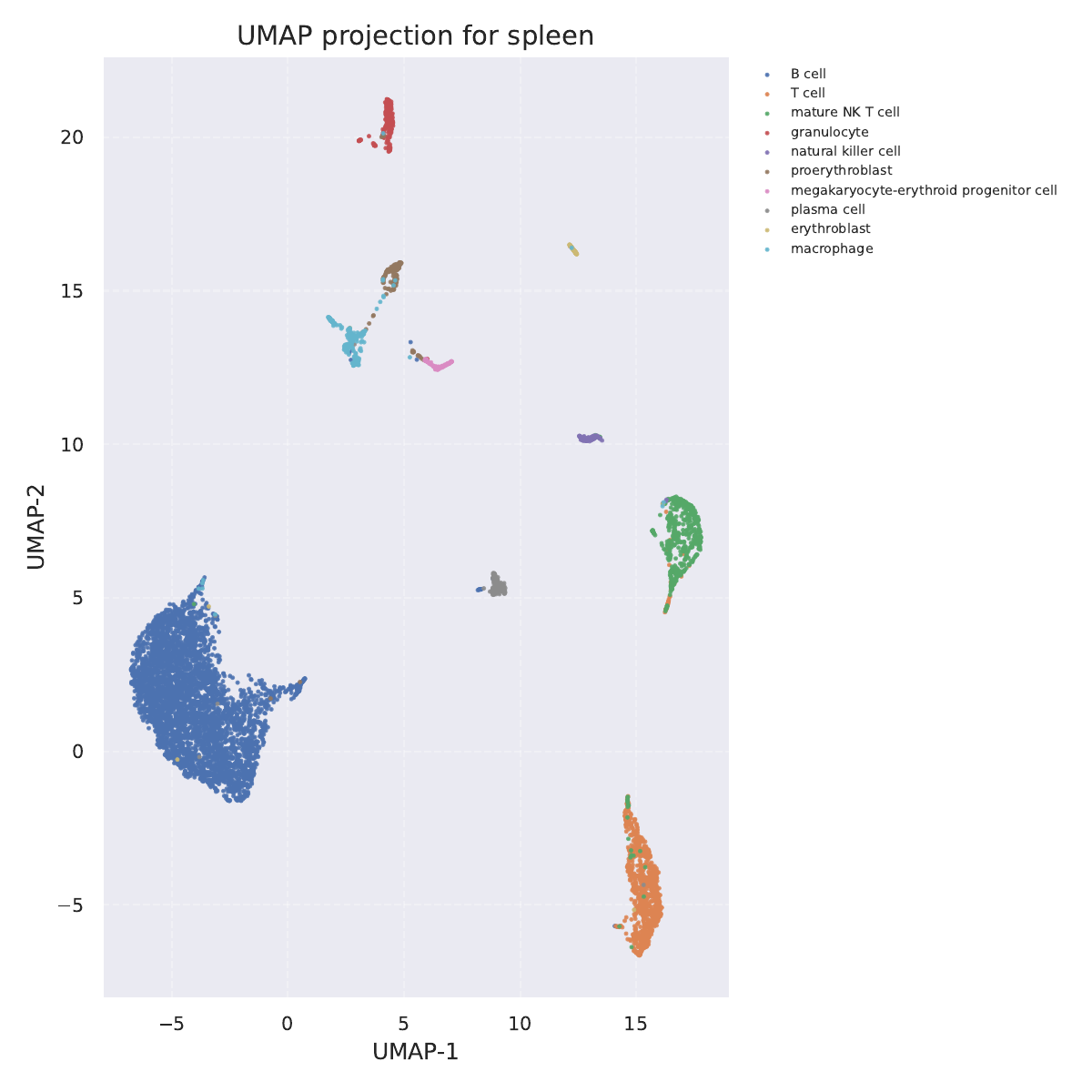}
        \caption{Mouse-Geneformer}
        \label{fig:umap_mouse_geneformer}
    \end{subfigure}
    \caption{UMAP visualizations of mouse spleen cells using Mix-Geneformer and Mouse-Geneformer.}
    \label{fig:spleen_umap_mouse_comparison}
\end{figure}

The fine-tuning and zero-shot cell-type classification accuracies for Mouse-Geneformer and the two Mix-Geneformer variants across different mouse tissues are summarized in Table~\ref{mouse-classification}. In the fine-tuning setting, both Mix-Geneformer and Mouse-Geneformer achieved high accuracy (over 95\%) in most organs, demonstrating robust performance in key tissues such as the brain, heart, and kidney. We observed that Mix-Geneformer outperformed Mouse-Geneformer in the mammary gland and thymus, while the accuracy was similar for both models in the tongue. In the zero-shot setting (Table~\ref{mouse-classification}), Mix-Geneformer without SimCSE-based contrastive learning (w/o CL) yielded the highest overall accuracy, showing improved performance over Mouse-Geneformer in the limb muscle, mammary gland, and spleen. These results indicate that removing contrastive learning may contribute to avoiding overfitting to species-specific characteristics, leading to more generalized representations and improved zero-shot classification of mouse cells. It's important to note that the inclusion of SimCSE-based contrastive learning (w/ CL) did not consistently result in better performance across both settings. While contrastive learning is generally beneficial for representation generalization, its effect on zero-shot performance appears to be complex and can lead to a decrease in accuracy if the learned representations are not well-aligned with the specific class boundaries of the evaluation dataset. Additionally, visual inspection of the UMAP embeddings for the ten annotated spleen cell types shows that the cluster centroids for Mouse-Geneformer and Mix-Geneformer align within approximately one UMAP unit along both axes. Major populations—such as B cells adjacent to plasma cells, T cells opposite granulocytes, and the erythroid lineage forming a continuous trajectory—maintain consistent spatial relationships in both embeddings Figure~\ref{fig:spleen_umap_mouse_comparison}. Cluster compactness, particularly for mature NK T cells and natural killer cells, is comparable between models, with no observable increase in dispersion in the Mix-Geneformer embedding. These visual concordances suggest that Mix-Geneformer preserves both the global topology and local density structure of the spleen feature manifold, reinforcing its ability to capture biologically meaningful representations on par with the species-specific Mouse-Geneformer. We also present the comparison of cell-type classification accuracies between Mix-Geneformer and previous models in Table~\ref{comparison}. As a result, Mix-Geneformer achieves the highest accuracy, demonstrating its effectiveness compared to conventional methods. 


\subsubsection{Cell-Type Classification in Human}
\begin{figure}[tb]
    \centering
    \begin{subfigure}[b]{0.48\linewidth}
        \centering
        \includegraphics[width=\linewidth]{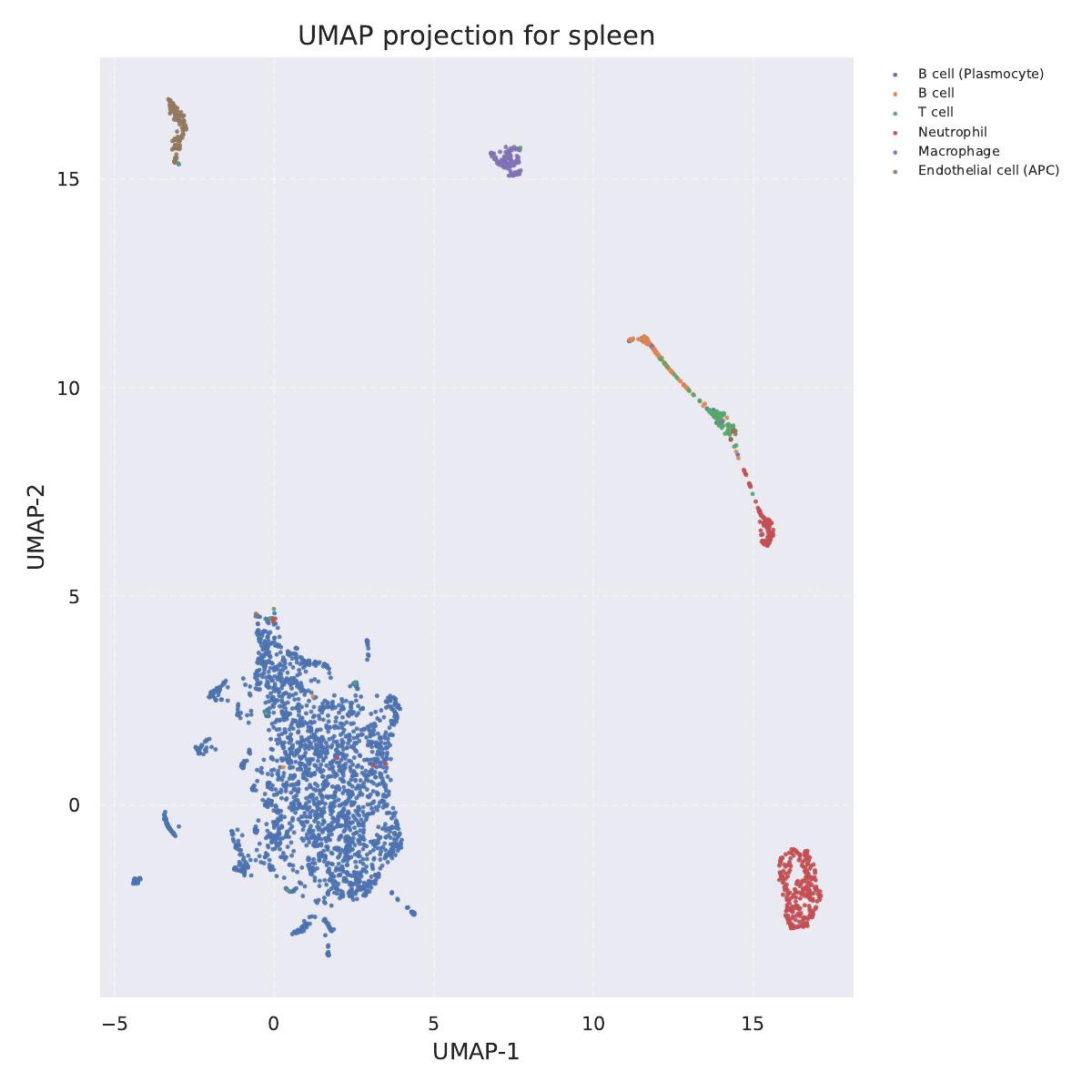}
        \caption{Mix-Geneformer}
        \label{fig:umap_mix_geneformer}
    \end{subfigure}
    \hfill
    \begin{subfigure}[b]{0.48\linewidth}
        \centering
        \includegraphics[width=\linewidth]{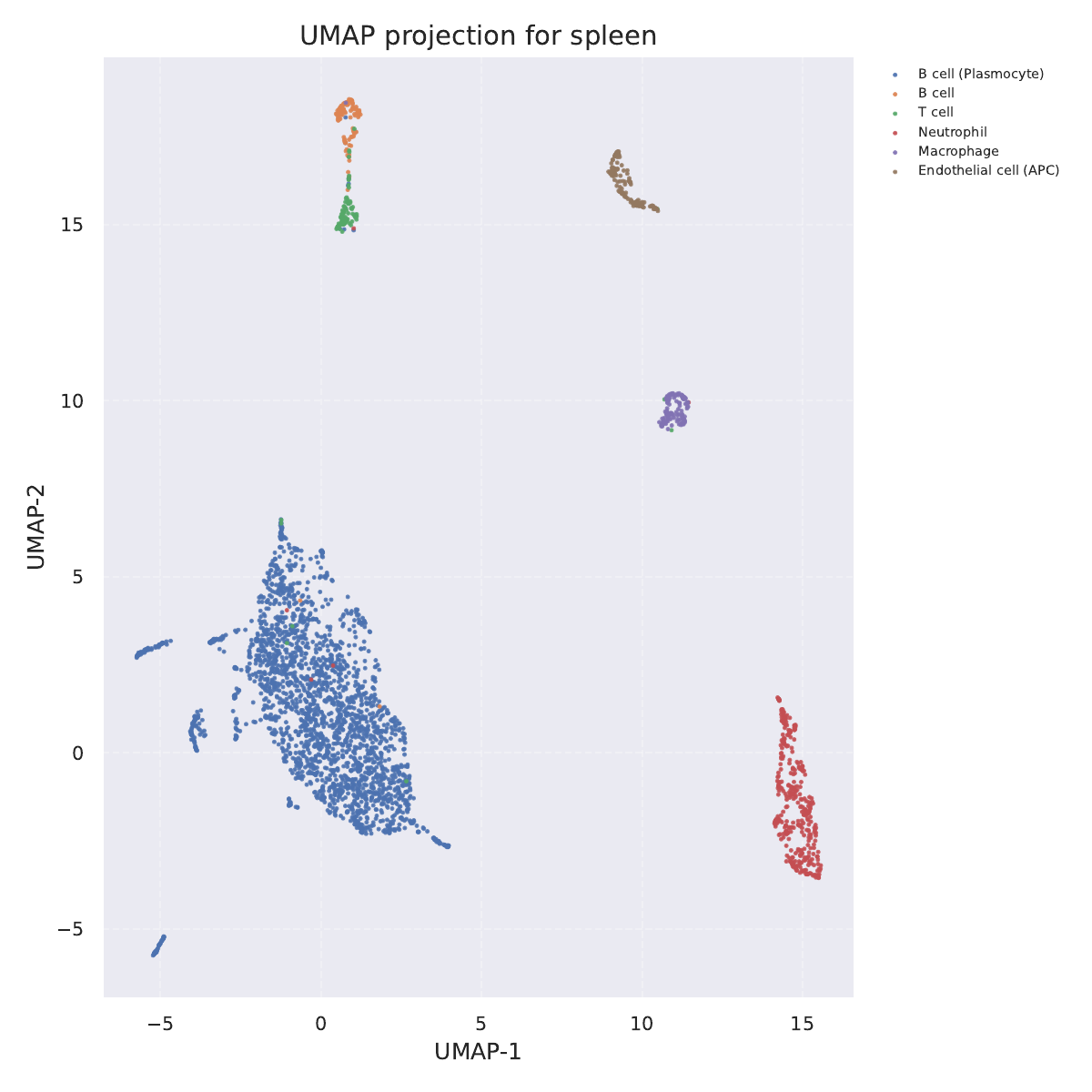}
        \caption{Human-Geneformer}
        \label{fig:umap_human_geneformer}
    \end{subfigure}
    \caption{UMAP visualization of human spleen cells using Mix-Geneformer and Human-Geneformer}
    \label{fig:umap_human_comparison}
\end{figure}

Table~\ref{human-classification} details the fine-tuning and zero-shot cell-type classification performance of Human-Geneformer and the two Mix-Geneformer variants on human tissues. During fine-tuning, both models demonstrated high accuracy across various human tissues, with Human-Geneformer exhibiting a slightly better overall performance, though the differences were minor. In the more challenging zero-shot scenario, Human-Geneformer generally achieved the highest accuracy. Interestingly, Mix-Geneformer (w/ CL) showed superior performance to Human-Geneformer in the placenta, highlighting its potential for cross-species generalization in specific contexts, while lagging in immune, kidney, and lung tissues. This trade-off likely reflects Human-Geneformer's inherent bias towards human data due to its training. As observed in the mouse experiments, the effect of SimCSE-based contrastive learning on zero-shot performance in human data was not uniform. While beneficial in some organs, it negatively affected others, possibly due to a mismatch between the learned generalized representations and the target dataset's specific cell-type distinctions. Furthermore, we examined the UMAP embeddings of the nine annotated human spleen cell types (Figure~\ref{fig:umap_human_comparison}).  Based on the UMAP projections in Figure~\ref{fig:umap_human_comparison}, we correctly observe that the centroids of corresponding cell‐type clusters for Mix‐Geneformer and Human‐Geneformer align within approximately one UMAP unit along both axes. Major populations—such as B cells adjacent to plasmocytes, T cells opposite neutrophils, and endothelial cells bordering macrophages—maintain consistent spatial relationships in both embeddings, and cluster compactness is comparable with no increased dispersion in the Mix‐Geneformer embedding.

\begin{table}[tb]
\caption{Cell-type classification accuracy for Mouse-Geneformer and Mix-Geneformer}
\label{mouse-classification}
\makebox[\textwidth][c]{%
  \begin{tabular}{c|c|c|c|c|c|c|c}
    \hline
    \textbf{Tissues}
      & \textbf{\makecell[c]{Cell\\Types}}
      & \textbf{\begin{tabular}[c]{@{}c@{}}Mouse-GF\\FT\end{tabular}}
      & \textbf{\begin{tabular}[c]{@{}c@{}}Mix-GF\\(w/ CL)\\FT\end{tabular}}
      & \textbf{\begin{tabular}[c]{@{}c@{}}Mix-GF\\(w/o CL)\\FT\end{tabular}}
      & \textbf{\begin{tabular}[c]{@{}c@{}}Mouse-GF\\zero-shot\end{tabular}}
      & \textbf{\begin{tabular}[c]{@{}c@{}}Mix-GF\\(w/ CL)\\zero-shot\end{tabular}}
      & \textbf{\begin{tabular}[c]{@{}c@{}}Mix-GF\\(w/o CL)\\zero-shot\end{tabular}} \\
    \hline
    brain       & 15  & 96.9 & 97.3 & 96.7 & 86.8 & 85.4 & 88.5 \\
    heart        & 11  & 97.8 & 97.7 & 97.2 & 89.1 & 88.1 & 88.8 \\
    kidney       & 18 & 94.9 & 95.6 & 94.4 & 87.3 & 85.6 & 88.2 \\
    large\_intestine       & 7 & 94.3 & 94.7 & 93.1 & 66.3 & 64.6 & 67.2 \\
    limb\_muscle & 9 & 99.5 & 99.6 & 99.4 & 96.3 & 94.7 & 97.2 \\
    mammary\_gland        & 7 & 99.0 & 99.1 & 99.1 & 97.1 & 96.2 & 97.3 \\
    spleen         & 10 & 98.7 & 98.7 & 98.3 & 94.2 & 91.2 & 95.8 \\
    thymus     & 6 & 97.0 & 97.5 & 97.2 & 84.4 & 76.9 & 89.3 \\
    tongue     & 3  & 94.9 & 94.8 & 93.6 & 83.2 & 81.6 & 84.4 \\
    \hline
  \end{tabular}
  }
\end{table}

\begin{table}[tb]
\caption{Cell-type classification accuracies of Mix-Geneformer and conventional models}
\label{comparison}
\makebox[\textwidth][c]{%
\begin{tabular}{c|c|c|c|c|c}
\hline
\textbf{Tissues} & \textbf{Cell Types} & \textbf{\begin{tabular}[c]{@{}c@{}}Mouse-\\ Geneformer\end{tabular}} & \textbf{Mix-Geneformer} & \textbf{scDeepsort} & \textbf{scVAE} \\ \hline
brain & 15 & 96.9 & 97.3 & 58.5 & 76.2 \\
heart & 11 & 97.8 & 97.7 & 79.6 & 79.4 \\
kidney & 18 & 94.9 & 95.6 & 57.6 & 56.2 \\
large\_intestine & 7 & 93.1 & 94.7 & 58.5 & 59.0 \\
limb\_muscle & 9 & 99.5 & 99.6 & 90.8 & 79.6 \\
mammary\_gland & 7 & 99.0 & 99.1 & 48.0 & 79.5 \\
spleen & 10 & 98.7 & 98.7 & 81.0 & 76.5 \\
thymus & 6 & 97.0 & 97.5 & 55.0 & 75.0 \\
tongue & 3 & 94.9 & 94.8 & 76.8 & 80.4 \\ \hline
\end{tabular}
}
\end{table}

\begin{table}[tb]
\caption{Cell-type classification accuracy for Human-Geneformer and Mix-Geneformer}
\label{human-classification}
\makebox[\textwidth][c]{%
  \begin{tabular}{c|c|c|c|c|c|c|c}
    \hline
    \textbf{Tissues}
      & \textbf{\makecell[c]{Cell\\Types}}
      & \textbf{\begin{tabular}[c]{@{}c@{}}Human-GF\\FT\end{tabular}}
      & \textbf{\begin{tabular}[c]{@{}c@{}}Mix-GF\\(w/ CL)\\FT\end{tabular}}
      & \textbf{\begin{tabular}[c]{@{}c@{}}Mix-GF\\(w/o CL)\\FT\end{tabular}}
      & \textbf{\begin{tabular}[c]{@{}c@{}}Human-GF\\zero-shot\end{tabular}}
      & \textbf{\begin{tabular}[c]{@{}c@{}}Mix-GF\\(w/ CL)\\zero-shot\end{tabular}}
      & \textbf{\begin{tabular}[c]{@{}c@{}}Mix-GF\\(w/o CL)\\zero-shot\end{tabular}} \\
    \hline
    spleen       & 6  & 98.9 & 98.3 & 97.9 & 78.5 & 78.8 & 78.5 \\
    brain        & 6  & 96.8 & 95.9 & 96.0 & 87.2 & 87.0 & 87.2 \\
    immune       & 10 & 94.4 & 93.6 & 91.1 & 42.3 & 38.4 & 39.6 \\
    kidney       & 15 & 92.8 & 92.2 & 89.7 & 45.5 & 38.4 & 37.3 \\
    large\_intestine & 16 & 92.7 & 91.5 & 89.7 & 43.7 & 40.1 & 40.5 \\
    liver        & 12 & 91.1 & 88.0 & 87.9 & 47.7 & 44.3 & 45.2 \\
    lung         & 16 & 93.4 & 92.6 & 91.5 & 50.1 & 39.7 & 40.4 \\
    pancreas     & 15 & 93.0 & 92.2 & 90.0 & 52.3 & 42.5 & 52.3 \\
    placenta     & 3  & 97.9 & 97.4 & 96.1 & 75.2 & 80.7 & 75.2 \\
    \hline
  \end{tabular}
  }
\end{table}

\subsection{Evaluation by In Silico Perturbation Experiments}

We conducted three in silico perturbation experiments to evaluate Mix-Geneformer’s ability to model disease-related gene perturbations in both human and mouse. Specifically, we simulated : (1) the shift of healthy mouse cells toward a diabetic kidney disease (DKD) state, (2) the shift of healthy mouse cells toward an autosomal dominant polycystic kidney disease (ADPKD) state, and (3) the shift of healthy human cells toward an Alzheimer’s disease state. The simulation results are summarized in Table~\ref{in-silico}, which lists for each disease model the top-ranking genes alongside their cosine\_shift values, ranks, and corresponding p-values.

In the DKD simulation, overexpression of genes previously implicated in disease progression—Spp1, Sgk1, Cfb, and Rock2—induced a pronounced shift in cellular state toward the DKD condition, demonstrating Mix-Geneformer’s ability to accurately identify key DKD-associated genes. Similarly, in the ADPKD simulation, deletion of the Umod gene, a known contributor to disease progression, caused a marked shift toward the ADPKD state, further supporting the model’s capacity to pinpoint disease-relevant genes. Finally, in the Alzheimer’s disease experiment using human cells, overexpression of S100B, HSPB1, and HSP90AA1—genes previously linked to Alzheimer’s in vivo—resulted in a robust transition toward the Alzheimer’s condition. Together, these findings confirm the effectiveness of Mix-Geneformer in both mouse and human in silico perturbation experiments.

Collectively, these results demonstrate that Mix-Geneformer effectively models in silico perturbations in both human and mouse, enabling the identification and prediction of disease-associated genes. This capability positions Mix-Geneformer as a promising tool for drug discovery and elucidating disease mechanisms.

\begin{table}[tb]
\caption{Top‐ranking genes identified by in silico perturbation across three disease models (DKD, ADPKD, Alzheimer’s), showing cosine\_shift scores and statistical significance.}
\label{in-silico}
\makebox[\textwidth][c]{%
\begin{tabular}{c|c|c|c|c}
\hline
\textbf{Gene\_name} & \textbf{disease} & \textbf{rank} & \textbf{cosine\_shift} & \textbf{p\_value} \\ \hline
Umod & ADPKD (Mouse) & 19 & 0.0237 & 2.54e-64 \\ \hline
Sgk1 & DKD (Mouse) & 19 & 0.0161 & 1.47e-04 \\
Cfb & DKD (Mouse) & 31 & 0.0132 & 1.72e-04 \\
Rock2 & DKD (Mouse) & 51 & 0.0082 & 1.60e-05 \\ \hline
HSPB1 & Alzheimer (Human) & 11 & 0.0544 & 1.58e-13 \\
S100B & Alzheimer (Human) & 20 & 0.0425 & 5.78e-05 \\
HSP90AA1 & Alzheimer (Human) & 112 & 0.0143 & 9.99e-07 \\ \hline
\end{tabular}
}
\end{table}

\section{Conclusion}

This study introduced Mix-Geneformer, a novel deep learning model for the integrated analysis of human and mouse gene expression data. By combining Masked Language Modeling (MLM) with SimCSE-based contrastive learning, Mix-Geneformer effectively captures both shared and species-specific gene representations. Evaluation experiments demonstrate that Mix-Geneformer achieves performance comparable to or exceeding existing species-specific models in cell-type classification and in silico perturbation tasks, validating its effectiveness. 

Notably, the in silico perturbation experiments successfully included human Alzheimer’s disease models, showcasing the model's ability to identify disease-associated genes in humans, beyond traditional mouse-only analysis. These results position Mix-Geneformer as a general-purpose model for unified human and mouse expression data analysis. The model's capacity to accelerate the translation of findings from mouse models to human contexts offers the potential to streamline translational research and reduce resource demands in drug discovery and disease analysis. 

This work highlights the promise of cross-species representation learning for advancing data-driven biological and medical research. Future work will focus on expanding the model to include other organisms and applying it to more complex disease areas like cancer and immune disorders, as well as exploring downstream applications such as disease prediction and drug target identification. 
\bibliographystyle{unsrt}  
\bibliography{references}  

\end{document}